\documentstyle[preprint,prd,aps]{revtex}
\begin{document}
\draft
\title{
Reaction-rate formula in out of equilibrium quantum field theory} 
\author{A. Ni\'{e}gawa, K. Okano, and H. Ozaki} 
\address{Department of Physics, Osaka City University, 
Sumiyoshi-ku, Osaka 558-8585, JAPAN\thanks{Electronic addresses: 
niegawa@sci.osaka-cu.ac.jp, okano@sci.osaka-cu.ac.jp, 
hozaki@sci.osaka-cu.ac.jp}}
\date{Received today}
\maketitle
\begin{abstract}
A complete derivation, from first principles, of the reaction-rate 
formula for a generic reaction taking place in an out of equilibrium 
quantum-field system is given. It is shown that the formula involves 
no finite-volume correction. Each term of the reaction-rate formula 
represents a set of physical processes that contribute to the 
reaction under consideration. 
\hspace*{1ex} 
\end{abstract} 
\narrowtext 
\setcounter{equation}{0}
\setcounter{section}{0}
\section{Introduction} 
\def\theequation{\mbox{\arabic{section}.\arabic{equation}}}
Ultrarelativistic heavy-ion-collision experiments at the BNL 
Relativistic Heavy Ion Collider (RHIC) and at the CERN Large Hadron 
Collider (LHC) will soon start in anticipation of producing 
quark-gluon plasma (QGP). Confirmation of the QGP formation is done 
through analyzing rates of various reactions taking place in a QGP. 
So far, the reaction-rate formula is derived for reactions taking 
place in the system in thermal and chemical equilibrium 
\cite{wel,nie,jac,niesin}. The actual QGP is, however, not in 
equilibrium 
but is an expanding nonequilibrium system. 

In this paper, as a generalization of \cite{wel,nie,jac,niesin}, we 
present a first-principles derivation of the reaction-probability 
formula for reactions occurring in a nonequilibrium system. We find 
that the formula involves no finite-volume corrections. We also find 
from the procedure of derivation that different contributions to the 
reaction-probability formula have clear physical interpretation, 
which is summarized as \lq\lq out-of-equilibrium cutting rules.'' 

In Sec.~II, we derive from first principles the formula for the 
transition probability of a generic reaction taking place in a 
nonequilibrium system. In Sec.~III, specializing to quasi\-uniform 
systems near equilibrium or non\-equilibrium quasistationary 
systems,\footnote{Framework for dealing with such systems is 
comprehensively discussed in \cite{chou}.} we further deduce the 
formula, finding that the formula is written in terms of the 
closed-time-path formalism of real-time thermal field theory 
\cite{chou}. In Sec.~IV, we present a calculational procedure 
of a generic reaction-probability formula obtained in Sec.~III. 
\setcounter{equation}{0} 
\setcounter{section}{1} 
\section{Nonequilibrium reaction-probability formula} 
\subsection{Preliminaries} 
\def\theequation{\mbox{\arabic{section}.\arabic{equation}}} 
The formalism presented in this paper can be applied to a broad 
class of theories including QCD (cf. the end of Sec.~III), but, for 
simplicity of presentation, we take a system of self-interacting, 
neutral scalars $\phi$'s with mass $m$ and $\lambda \phi^4$ 
interaction. The system is inside a cube with volume $V = L^3$. 
Employing the periodic boundary conditions, we label the 
single-particle basis by its momentum ${\bf p_{\bf k}} = 2\pi {\bf 
k} / L$, $k_j = 0,\pm 1, \pm 2, \cdots, \pm \infty \; (j = 1, 2, 
3)$. 

Physically interesting reactions are of the following generic type, 
\begin{equation} 
\{ A \} + \mbox{nonequilibrium system} \to \{ B \} + \mbox{anything} 
\, . 
\label{jyo} 
\end{equation} 
Here $\{ A \}$ and $\{ B \}$ designate group of particles, which are 
different from $\phi$. Examples are highly virtual particles, heavy 
particles, and particles interacting weakly with $\phi$'s. 
Generalization to more general process, where among $\{ A \}$ 
and/or $\{ B \}$ are $\phi$'s, is straightforward (cf., 
\cite{niesin}). For definiteness, let us assume that $\{ A \}$ 
consists of $l$ $\Phi$'s and $\{ B \}$ consists of $l'$ $\Phi$'s. 
Here $\Phi$ is a heavy neutral scalar of mass $M$, so that $\Phi$ is 
absent in the system. For simplicity of presentation, we assume a 
$\Phi$-$\phi$ coupling to be of the form $- g \Phi\phi^n / n!$ $(n 
\geq 2)$. 

The transition or reaction probability ${\cal P}$ of the process 
(\ref{jyo}) is written as 
\begin{mathletters} 
\label{maru} 
\begin{eqnarray}
{\cal P} & = & {\cal N} / {\cal D} \:, \eqnum{2.2a} 
\label{P} \\ 
{\cal N} & \equiv & \sum_{\{ {\bf k} \}} \sum_{\{ n_{\bf k} \}} \; 
\sum_{\{ m_{\bf k} \}} \; \sum_{\{ n_{\bf k}' \}} \langle \{ A \}; 
\{ m_{\bf k} \} | \, S^\dagger \, | \{ n_{\bf k}' \}; \{ B \} 
\rangle \nonumber \\ 
&& \times \langle \{ B \}; \{ n_{\bf k}' \} | \, S \, | \{ 
n_{\bf k} \}; \{ A \} \rangle \, \langle \{ n_{\bf k} \} | \, \rho \, 
| \{ m_{\bf k} \} \rangle \, {\bf S} \, , \nonumber \\ 
\eqnum{2.2b} 
\label{cal-N} \\ 
{\cal D} & \equiv & \sum_{\{ {\bf k} \}} \sum_{\{ n_{\bf k} \}} \; 
\sum_{\{ m_{\bf k} \}} \; \sum_{\{ n_{\bf k}' \}} \langle \{ m_{\bf 
k} \} | \, S^\dagger \, | \{ n_{\bf k}' \} \rangle \nonumber \\ 
&& \times \langle \{ n_{\bf k}' \} | \, S \, | \{ n_{\bf k} \} 
\rangle \, \langle \{ n_{\bf k} \} | \, \rho \, | \{ m_{\bf k} \} 
\rangle \, {\bf S} \, . 
\eqnum{2.2c} 
\label{cal-D} 
\end{eqnarray}
\end{mathletters} 
Here ${\bf S}$ is the symmetry factor \cite{BD}, $\rho$ is the 
density matrix, and $\langle \{ B \}; \{ n_{\bf k}' \}| \, S \, 
| \{ n_{\bf k} \}; \{ A \} \rangle$ is a $S$-matrix element 
of the {\em vacuum-theory} {\em process}, 
\[ 
\{ A \} + \{ n_{\bf k} \} \to \{ B \} + \{ 
n_{\bf k}' \} \, , 
\] 
where $\{ n_{\bf k} \}$ denotes the group of $\phi$'s, which 
consists of the number $n_{\bf k}$ of $\phi_{\bf k}$ ($\phi$ in a 
mode ${\bf k}\; )$. In Eqs.~(\ref{maru}), $\sum_{\{ {\bf k} \}}$ 
denotes summation over momentum/momenta of $\phi$/$\phi$'s in the 
final state $| \{ n_{\bf k}' \} ; \{ B \} \rangle$, and of 
$\phi$/$\phi$'s in the \lq\lq two'' initial states $| \{ m_{\bf k} 
\} ; \{ A \} \} \rangle$ and $| \{ n_{\bf k} \} ; \{ A \} \rangle$. 
(Among the final states $| \{ n_{\bf k}'\} ; \{ B \} \rangle$, is 
$| 0 ; \{ B \} \rangle$. This is also the case for \lq\lq two'' 
initial states.) Note that the perturbation series for ${\cal D}$ 
starts from 1, 
\[ 
{\cal D} = 1 + \cdots \, . 
\] 
It is to be noted that $\{ A \}$ and $\{ B \}$ in $\langle S 
\rangle$, which we write $\{A, \, B \}_S$, are not necessarily 
involved in one connected part of $\langle S \rangle$. This is also 
the case for $\{A, \, B \}_{S^\dagger}$. We assume that, in $W 
\equiv \langle S^\dagger \rangle \langle S \rangle$, $\{A, \, B 
\}_S$ and $\{A, \, B \}_{S^\dagger}$ are involved in one connected 
part $W_c$ $(\in W)$. Then, $W$ consists, in general, of $W_c$ and 
other parts which are disconnected with $W_c$ and include only 
$\phi$'s. Generalization to other cases is straightforward 
\cite{niesin}. 
It should be remarked on the form of $\rho$ in Eqs.~(\ref{maru}). 
Let us recall the following two facts. On the one hand, the 
statistical ensemble is defined by the density matrix at the very 
initial time $t_i$ ($\sim - \infty$). On the other hand, in 
constructing perturbative framework, an adiabatic switching off of 
the interaction is required \cite{lan,jac}. Then, $\rho$ in 
Eqs.~(\ref{maru}) is a functional of the in-field $\phi_{in} (t_i, 
{\bf x})$ that constitutes the basis of perturbation theory. 

As will be seen below, diagrammatic analysis shows that ${\cal N}$, 
Eq.~(\ref{cal-N}), takes the form, 
\begin{equation} 
{\cal N} = {\cal N}_{\mbox{\scriptsize{con}}} \, {\cal D} \, , 
\label{hahaha} 
\end{equation} 
where ${\cal N}_{\mbox{\scriptsize{con}}}$ corresponds to a 
connected diagram and ${\cal D}$ is as in Eq.~(\ref{cal-D}). Then, 
we have 
\[ 
{\cal P} = {\cal N}_{\mbox{\scriptsize{con}}} \, . 
\] 

The $S$-matrix element in vacuum theory is obtained through an 
application of the reduction formula \cite{nie,niesin}: 
\widetext 
\begin{eqnarray} 
&& \langle \{ B \}; \{ n_{\bf k}' \} | \, S \, | {\{ n_{\bf k} \}}; 
\{ A \} \rangle \nonumber \\ 
& & \mbox{\hspace*{4ex}} = \prod_{j=1}^l \left( i K_{j, \Phi_j} 
\right) \prod_{m = 1}^{l'} \left( iK_{m, \Phi_m}^* \right) 
\Big\langle 0 \Bigm| \prod_{\bf k} T \left[ \left\{ \sum_{i_{\bf k} 
= 0}^{n_{\bf k} } \sum_{i'_{\bf k} = 0}^{n'_{\bf k}} \delta(n_{\bf 
k} - i_{\bf k} \; ; \; n'_{\bf k} - {i'_{\bf k}}) N^{n_{\bf k} \, 
n_{\bf k}{}'}_{i_{\bf k} \, i'_{\bf k}} \right. \right. \nonumber \\ 
& & \left. \left. \mbox{\hspace*{7ex}} \times \prod_{n'=1}^{i'_{\bf 
k}} \left(iK_{{\bf k},n'}^* \right) \prod_{n = 1}^{i_{\bf k}} \left( 
i K_{{\bf k},n} \right) \prod_{n'=1}^{i'_{\bf k}} \phi_{n'} 
\prod_{n = 1}^{i_{\bf k}} \phi_n \right\} \prod_{j = 1}^l \Phi_j 
\prod_{m = 1}^{l'} \Phi_m \right] \Bigm| 0 \Big\rangle \, , 
\label{S} 
\end{eqnarray}
\narrowtext 

\noindent where $T$ is the time-ordering symbol and 
\begin{equation} 
N^{n_{\bf k} n'_{\bf k}}_{i_{\bf k} \, i'_{\bf k}} \equiv 
\left\{ \left(
\begin{array}{c}
{n'_{\bf k}} \\ 
{i'_{\bf k}} 
\end{array}  \right)
\left(
\begin{array}{c}
n_{{\bf k}} \\ 
i_{\bf k}
\end{array} \right) 
\frac{1}{i_{\bf k}'! \; i_{\bf k}!} \right\}^{1/2}. 
\label{N-def} 
\end{equation} 
In Eq.~(\ref{S}) $\delta(\cdots \, ; \, \cdots)$ denotes the 
Kronecker's $\delta$-sym\-bol and 
\begin{eqnarray} 
K_{{\bf k},n} \cdots \phi_{n} & \equiv & {1 \over 
{\sqrt{Z_{\phi}}}} \int \! d^4 x \, f_{{\bf p}_{\bf k}} (x) \, 
(\Box + m^2) \, \cdots \phi (x) \, , \nonumber \\ 
K_{j, \, \Phi_j} \cdots \Phi_j & \equiv & {1 \over \sqrt{Z_{\Phi}}} 
\int d^4 x \, F_j (x) (\Box + M^2) \, \cdots  \Phi (x) \, , 
\nonumber \\ 
K^*_{m, \, \Phi_m} \cdots \Phi_m & \equiv & {1 \over 
\sqrt{Z_{\Phi}}} \int d^4 x \, G^*_m (x) (\Box + M^2) \, \cdots  
\Phi (x) \, . \nonumber \\ 
\label{K} 
\end{eqnarray} 
Here 
\[ 
f_{{\bf p}_{\bf k}} (x) = \frac{1}{\sqrt{2 E_{\bf k} V}} \, \, e^{- 
i P_{\bf k} \cdot x} \, , \;\;\;\;\;\;\; (E_{\bf k} = \sqrt{p_{\bf 
k}^2 + m^2} \, ) \, , 
\] 
with $P_{\bf k}^\mu \equiv (E_{\bf k}, {\bf p}_{\bf k})$ and $F_j 
(x)$ [$G^*_m (x)$] the wave function of $j$th $\Phi$ $(\in \{ A 
\})$ [$m$th $\Phi$ $(\in \{ B \})$]. $Z$'s in Eq.~(\ref{K}) are the 
wave-function renormalization constants. It is to be noted that, in 
Eq.~(\ref{S}), among $n_{\bf k}$ ($n'_{\bf k}$) of ${\phi_{\bf 
k}}$'s in the initial (final) state, $i_{\bf k}$ ($i'_{\bf k}$) of 
${\phi_{\bf k}}$'s are absorbed in (emitted from) the $i_{\bf k}$ 
($i'_{\bf k}$) vertices in $S$. Remaining $n_{\bf k} - i_{\bf k}$ ($ 
= n'_{\bf k} - i'_{\bf k}$) of ${\phi_{\bf k}}$'s  are merely 
spectators, which reflects only on the statistical factor in ${\cal 
F}_i$ in Eq.~(\ref{owayo}) below. 

$\langle S \rangle$ in ${\cal D}$ in Eq.~(\ref{cal-D}) is given by a 
similar expression to Eq.~(\ref{S}), where factors related to the 
$\Phi$ fields are deleted. 

>From the form for $\langle S \rangle$, Eq.~(\ref{S}), we see that 
the permutation of $\phi_n$ ($n = 1, \cdots, i_{\bf k}$) and the 
permutation of $\phi_{n'}$ ($n' = 1, \cdots, i_{\bf k}'$) give the 
same Feynman diagram (in vacuum theory), and then $i_{\bf k} ! 
i_{\bf k}' !$ same diagrams emerge. Taking this fact into account, 
we may write (\ref{S}) in the form, 
\widetext 
\begin{eqnarray} 
&& \langle \{ B \}; \{ n_{\bf k}' \} | \, S \, | \{ n_{\bf k} \}; 
\{ A \} \rangle \nonumber \\ 
&& \mbox{\hspace*{4ex}} = \left( \prod_{j = 1}^l \int d^{\, 4} x_j 
F_j (x_j) \right) \left( \prod_{m = 1}^{l'} \int d^{\, 4} y_m G^*_m 
(y_m) \right) \nonumber \\ 
&& \mbox{\hspace*{7ex}} \times \sum_{\{ i_{\bf k} \}} 
\left[ \prod_{\bf k} N^{n_{\bf k} \, n_{\bf k}'}_{i_{\bf k} \, 
i'_{\bf k}} i_{\bf k} ! i_{\bf k}' ! \left( \prod_{j = 1}^{i_{\bf 
k}} \int d^{\, 4} \xi_{{\bf k} \, j} f_{{\bf p}_{\bf k}} 
(\xi_{{\bf k} \, j}) \right) \left( \prod_{j = 1}^{i_{\bf k}'} \int 
d^{\, 4} \zeta_{{\bf k} \, j} f_{{\bf p}_{\bf k}}^* (\zeta_{{\bf k} 
\, j}) \right) 
\right] \nonumber \\ 
&& \mbox{\hspace*{7ex}} \times {\cal A} ( \{ y \}, \{ \zeta \}; \{ 
\xi \}, \{ x \} ) \, , 
\label{2.10} 
\end{eqnarray} 
where $i_{\bf k}' = n_{\bf k}' - n_{\bf k} + i_{\bf k}$ and ${\cal 
A}$ is the truncated Green function in configuration space (in 
vacuum theory), and, e.g., $\{ y \}$ collectively denotes $y_1, y_2, 
\cdots, y_{l'} $. 

Among the Feynman diagrams for ${\cal A}$, are some diagrams, in 
which some $\xi$'s ($\in \{ \xi \}$) [$\zeta$'s ($\in \{ \zeta 
\}$)] coincide with $x$'s ($\in \{ x \}$) and/or $y$'s ($\in \{ y 
\}$) and/or $\zeta$'s ($\in \{ \zeta \}$) [$\xi$'s ($\in \{ \xi 
\}$)]. In such cases, ${\cal A}$ is understood to include 
corresponding $\delta$-functions, e.g., $\delta^{\, 4} (\xi_{{\bf 
k} \, j} - x_i)$. 

The expression for $\langle S^\dagger \rangle$, the complex 
conjugate of $\langle S \rangle$, is obtained by taking the complex 
conjugate of Eq.~(\ref{S}) or Eq.~(\ref{2.10}), where we make the 
substitution (cf. Eqs.~(\ref{cal-N}) and (\ref{cal-D})), 
\[ 
n_{\bf k} \to m_{\bf k}, \, \;\;\;\;\;\; n_{\bf k}' \to m_{\bf 
k}' \left( = n_{\bf k}' \right) \, \;\;\;\;\;\; i_{\bf k} \to 
j_{\bf k}, \, \;\;\;\;\;\; i'_{\bf k} \to j'_{\bf k} . 
\] 
This applies also to the expression for $\langle S^\dagger \rangle$ 
in Eq.~(\ref{cal-D}). 

Substitution of $W = \langle S^\dagger \rangle \langle S \rangle$ 
into Eq.~(\ref{cal-N}) yields, with obvious notation, 
\begin{eqnarray} 
{\cal N} & = & \left( \prod_{j = 1}^{l} \int d^{\, 4} x_j \, d^{\, 
4} x_j' \, F_j (x_j) F^*_j (x_j') \right) \left( \prod_{m = 1}^{l'} 
\int d^{\, 4} y_m \, d^{\, 4} y_m' \, G^*_m (y_m) G_m (y_m') \right) 
\nonumber \\ 
&& \times \sum_{\{ {\bf k} \}} \sum_{\{ i_{\bf k} \}} \sum_{\{ 
j_{\bf k} \}} \sum_{\{ i_{\bf k}' \}} \sum_{\{ j_{\bf k}' \}} \left[ 
\prod_{\bf k} \left( \prod_{j = 1}^{i_{\bf k}} \int d^{\, 4} 
\xi_{{\bf k} \, j} f_{{\bf p}_{\bf k}} (\xi_{{\bf k} \, j}) \right) 
\left( \prod_{j = 1}^{i_{\bf k}'} \int d^{\, 4} \zeta_{{\bf k} \, j} 
f_{{\bf p}_{\bf k}}^*(\zeta_{{\bf k} \, j}) \right) \right. 
\nonumber \\ 
&& \left. \times \left( \prod_{j = 1}^{j_{\bf k}} \int d^{\, 4} 
\xi'_{{\bf k} \, j} f^*_{{\bf p}_{\bf k}} (\xi'_{{\bf k} \, j}) 
\right) \left( \prod_{j = 1}^{j_{\bf k}'} \int d^{\, 4} \zeta'_{{\bf 
k} \, j} f_{{\bf p}_{\bf k}} (\zeta'_{{\bf k} \, j}) \right) 
\right] \nonumber \\ 
&& \times {\cal S} \, {\cal W} ( 
\{ x' \}, \{ \xi' \}; \{ \zeta' \}, \{ y' \} : \{ y \}, \{ \zeta \}; 
\{ \xi \}, \{ x \} ) \, {\bf S} \, . 
\label{N3} 
\end{eqnarray} 
Here $i_{\bf k}' = n_{\bf k}' - n_{\bf k} + i_{\bf k}$, $j_{\bf k}' 
= n_{\bf k}' - m_{\bf k} + j_{\bf k}$, ${\cal W} = {\cal A}^* {\cal 
A}$, and 
\begin{equation} 
{\cal S} \equiv \sum_{\{ n_{\bf k} \}} \left( \prod_{\bf k} 
N_{j_{\bf k} j_{\bf k}'}^{m_{\bf k} n_{\bf k}'} N_{i_{\bf k} i_{\bf 
k}'}^{n_{\bf k} n_{\bf k}'} i_{\bf k} ! i_{\bf k}' ! j_{\bf k} ! 
j_{\bf k}' ! \right) \langle \{ n_{\bf k} \} \mid \, \rho \, | \{ 
m_{\bf k} \} \rangle \, . 
\label{sin} 
\end{equation} 
\subsection{Statistical factor ${\cal S}$} 
Here, it is convenient to introduce creation and annihilation 
operators, $a_{{\bf p}_{\bf k}}^\dagger$ and $a_{{\bf p}_{\bf k}}$, 
which satisfy $[a_{{\bf p}_{\bf k}}, a_{{\bf p}_{{\bf k}'}}^\dagger 
] = \delta_{{\bf k}, {\bf k}'}$ and $[a_{{\bf p}_{\bf k}}, a_{{\bf 
p}_{{\bf k}'}}] = 0$. A Fock space ${\cal F}$ is constructed on $| 0 
\rangle$, which is defined by $a_{{\bf p}_{{\bf k}}} | 0 \rangle = 
0$. For the vector $| \;\;\; \rangle$ $(\in {\cal F})$ that 
satisfies $a_{{\bf p}_{{\bf k}}}^\dagger a_{{\bf p}_{{\bf k}}} | 
\;\;\; \rangle = n_{{\bf p}_{{\bf k}}} | \;\;\; \rangle$ ($n_{{\bf 
p}_{\bf k}} = 0, 1, 2, \cdots$), we use the same notation as in 
Eq.~(\ref{sin}), $| \{ n_{\bf k} \} \rangle$, since no confusion 
arises. A key observation here is that, using the form 
(\ref{N-def}), one can easily show that ${\cal S}$, Eq.~(\ref{sin}), 
may be represented as 
\begin{eqnarray} 
{\cal S} & = & \sum_{\{ n_{\bf k} \}} \Big\langle \{ m_{\bf k} \} 
\Bigm| \left( \prod_{l = 1}^j a^\dagger_{{\bf p}_l'} \right) \left( 
\prod_{l = 1}^{j'} a_{{\bf q}_l'} \right) \left( \prod_{l = 
1}^{i'} a^\dagger_{{\bf q}_l} \right) \left( \prod_{l = 1}^i 
a_{{\bf p}_l} \right) \Bigm| \{ n_{\bf k} \} \Big\rangle 
\Big\langle \{ n_{\bf k} \} \Bigm| \, \rho \, \Bigm| \{ m_{\bf k} 
\} \Big\rangle \nonumber \\ 
& \equiv & \Big\langle \left( \prod_{l = 1}^j a^\dagger_{{\bf 
p}_l'} \right) \left( \prod_{l = 1}^{j'} a_{{\bf q}_l'} \right) 
\left( \prod_{l = 1}^{i'} a^\dagger_{{\bf q}_l} \right) \left( 
\prod_{l = 1}^i a_{{\bf p}_l} \right) \Big\rangle \, , 
\label{sin1} 
\end{eqnarray} 
\narrowtext 

\noindent 
where we write 
\[ 
\{ {\bf p}_1, \, \cdots \, , {\bf p}_i \} = \{ \, \cdots \, , 
\underbrace{{\bf p}_{\bf k}, \, \cdots \, , {\bf p}_{\bf k}}_{i_{\bf 
k}}, \, \cdots \, \} \, , 
\] 
and then $i = \sum_{\bf k} i_{\bf k}$. Similarly, $i' = \sum_{\bf k} 
i_{\bf k}'$, $j = \sum_{\bf k} j_{\bf k}$, and $j' = \sum_{\bf k} 
j_{\bf k}'$. Note that $\langle \{ 
n_{\bf k} \}|$ and $| \{ m_{\bf k} \} \rangle$, in between which 
$\rho$ is sandwiched, are as in Eqs.~(\ref{maru}) and (\ref{sin}). 

Let us write ${\cal S}$, for short, as ${\cal S} = \langle b_1 b_2 
\, \cdots \, b_N \rangle$ $(N = i + j + i' + j')$. Let $l_1, \, 
\cdots \, , l_m$ be a solution in positive integers of 
\begin{equation} 
\sum_{j = 1}^m l_j = N \;\;\;\;\;\; (1 \leq m \leq N) \, . 
\label{seig} 
\end{equation} 
Pick out $l_1$ $b$'s out of $b_1, b_2, \, \cdots \, , b_N$ and pick 
out $l_2$ $b$'s out of remaining $b$'s, and so on, to make $m$ 
groups, 
\begin{equation} 
\left\{ b_1 \, \cdots \, b_{i_{l_1}} \right\} \left\{ b_{i_{l_1 + 1}} 
\, \cdots \, b_{i_{l_1 + l_2}} \right\} \cdot \cdot \cdot \left\{ 
b_{i_{N - {l_m} + 1}} \, \cdots \, b_{i_N} \right\} \, , 
\label{gr} 
\end{equation} 
where $1 < i_{l_1 + 1} < i_{l_1 + l_2 + 1} < \cdots < i_{N - {l_m} + 
1} \leq N$. In Eq.~(\ref{gr}), let $b_l$ and $b_{l'}$ are in between 
one set of curly brackets. Then, if $l < l'$, $b_l$ is located at 
the left of $b_{l'}$ and vice versa. 
We are now in a position to write 
\begin{eqnarray} 
&& {\cal S} (b_1 \, \cdots \, b_N) \nonumber \\ 
 && \mbox{\hspace*{4ex}} = \sum_{m = 1}^N 
\sum_{l\mbox{'\scriptsize{s}}} \sum_{\mbox{\scriptsize{gr}}} {\cal 
S}_c ( b_1 \, \cdots \, b_{i_{l_1}}) \nonumber \\ 
&& \mbox{\hspace*{7ex}} \times {\cal S}_c (b_{i_{l_1 + 1}} \, 
\cdots \, b_{i_{l_1 + l_2}} ) \cdots {\cal S}_c (b_{i_{N - {l_m} + 
1}} \, \cdots \, b_{i_N} ) \, . \nonumber \\ 
\label{ee} 
\end{eqnarray} 
Here, the second summation $\sum_{l\mbox{'\scriptsize{s}}}$ runs 
over all solutions in integers of Eq.~(\ref{seig}) and the third 
summation $\sum_{ \mbox{\scriptsize{gr}} }$ runs over all ways of 
making $m$ groups as in Eq.~(\ref{gr}). From Eq.~(\ref{ee}), ${\cal 
S}_c$ is determined iteratively. For example, 
\begin{eqnarray*} 
{\cal S}_c (b_1 b_2) & = & {\cal S} (b_1 b_2) - {\cal S} (b_1) 
{\cal S} (b_2) \\ 
{\cal S}_c (b_1 b_2 b_3) & = & {\cal S} (b_1 b_2 b_3) 
- {\cal S}_c (b_1 b_2) 
{\cal S} (b_3) 
- {\cal S}_c (b_1 b_3) 
{\cal S} (b_2) \\ 
&& - {\cal S} (b_1)  {\cal S}_c (b_2 b_3) 
- {\cal S} (b_1) {\cal S} (b_2) {\cal S} (b_3) 
\, . 
\end{eqnarray*} 
Thus, we have, with obvious notation, 
\begin{equation} 
{\cal S} = \sum_{m = 1}^{i + j + i' + j'} 
\sum_{l\mbox{'\scriptsize{s}}} \sum_{\mbox{\scriptsize{gr}}} 
{\cal S}_c (\cdots) {\cal S}_c (\cdots) \, \cdots \, {\cal S}_c 
(\cdots) \, . 
\label{haya} 
\end{equation} 
In the case of equilibrium system, all but $\langle a_{\bf p} 
a^\dagger_{\bf q} \rangle$ and $\langle a^\dagger_{\bf q} a_{\bf p} 
\rangle$ vanish. From the definition of ${\cal S}_c$, it is not 
difficult to show that, for $N \geq 3$, 
\begin{equation} 
{\cal S}_c (b_{j_1} b_{j_2} \, \cdots \, b_{j_l}) = 
{\cal S}_c (\, : b_{j_1} b_{j_2} \, \cdots \, b_{j_l} : \, ) \, , 
\label{ka} 
\end{equation} 
where \lq $: \cdots :$' indicates to take the normal ordering with 
respect to the creation and annihilation operators. 
\subsection{Reaction-probability formula} 
\widetext 
Now, ${\cal N}$ in Eq.~(\ref{N3}) may be written as  
\begin{eqnarray} 
{\cal N} & = & \left( \prod_{j = 1}^{l} \int d^{\, 4} x_j \, d^{\, 
4} x_j' F_j (x_j) F^*_j (x_j') \right) \left( \prod_{j = 1}^{l'} 
\int d^{\, 4} y_j \, d^{\, 4} y_j' G_j^* (y_j) G_j (y_j') \right) 
\nonumber \\ 
&& \times \sum_{i, \, j, \, i', \, j'} \left( \prod_{j = 1}^{i} \int 
d^{\, 4} \xi_j \sum_{{\bf p}_j} \frac{1}{\sqrt{2 E_{p_j} V}} e^{- i 
P_j \cdot \xi_j} \right) \left( \prod_{j = 1}^{i'} \int d^{\, 4} 
\zeta_j \sum_{{\bf q}_j} \frac{1}{\sqrt{2 E_{q_j} V}} e^{i Q_j \cdot 
\zeta_j} \right) \nonumber \\ 
&& \times \left( \prod_{l = 1}^{j} \int d^{\, 4} \xi_j' \sum_{{\bf 
p}_j'} \frac{1}{\sqrt{2 E_{p_j'} V}} e^{i P_j' \cdot \xi_j'} \right) 
\left( \prod_{l = 1}^{j'} \int d^{\, 4} \zeta_j' \sum_{{\bf q}_j'} 
\frac{1}{\sqrt{2 E_{q_j'} V}} e^{- i Q_j' \cdot \zeta_j'} \right) 
\nonumber \\    
&& \times {\cal S} \, {\cal W} (\{ x' \}, \{ \xi' \}; \{ \zeta' \}, 
\{ y' \} : \{ y \}, \{ \zeta \}; \{ \xi \}, \{ x \}) \, {\bf S} \, 
. 
\label{yatto} 
\end{eqnarray} 
Carrying out the integration over $\xi$'s, $\zeta$'s, $\xi'$'s, 
$\zeta'$'s and the internal spacetime vertex points, which are 
included in ${\cal W}$, we obtain, with obvious notation, 
\begin{eqnarray} 
{\cal N} & = & \left( \prod_{j = 1}^{l} \int d^{\, 4} x_j \, d^{\, 
4} x_j' F_j (x_j) F^*_j (x_j') \right) \left( \prod_{j = 1}^{l'} 
\int d^{\, 4} y_j \, d^{\, 4} y_j' G_j^* (y_j) G_j (y_j') \right) 
\nonumber \\ 
&& \times \sum_{i, \, j, \, i', \, j'} \left( \prod_{j = 1}^{i} 
\sum_{{\bf p}_j} \frac{1}{\sqrt{2 E_{{p}_j} V}} \right) \left( 
\prod_{j = 1}^{i'} \sum_{{\bf q}_j} \frac{1}{\sqrt{2 E_{{q}_j} V}} 
\right) \left( \prod_{l = 1}^{j} \sum_{{\bf p}_j'} \frac{1}{\sqrt{2 
E_{{p}_j'} V}} \right) \left( \prod_{l = 1}^{j'} \sum_{{\bf q}_j'} 
\frac{1}{\sqrt{2 E_{{q}_j'} V}} \right) \nonumber \\    
&& \times {\cal S} \, {\cal W} (\{ x' \}, \{ {\bf p}' \}; \{ 
{\bf q}' \}, \{ y' \} : \{ y \}, \{ {\bf q} \}; \{ {\bf p} \}, 
\{ x \}) \, {\bf S} \, . 
\label{yatto2} 
\end{eqnarray} 
Let us Fourier transform the wave functions $F_j (x)$, $G_j (x)$ 
\begin{eqnarray} 
F_j (x) = \int d {\bf r}_j \, e^{- i R_j \cdot (x - X_c)} 
\tilde{F}_j ({\bf r}_j) \, , \nonumber \\ 
G_j (x) = \int d {\bf r}_j \, e^{- i R_j \cdot (x - X_c)} 
\tilde{G}_j ({\bf r}_j) \, , 
\label{hadou} 
\end{eqnarray} 
where $R^\mu_j = (E_j, {\bf r}_j)$ with $E_j = \sqrt{r_j^2 + M^2}$. 
In Eq.~(\ref{hadou}), ${\bf X}_c$ of $X^\mu_c = (X_{c 0}, {\bf 
X}_c)$ is the space point, around which $\Phi$'s are localized and 
$X_{c 0}$ is the time, around which the reaction takes place. In 
general, $\tilde{F}_j$ and $\tilde{G}_j$ also depend on $X_c$. 

Substituting (\ref{hadou}) into Eq.~(\ref{yatto2}) and carrying out 
the integration over $x_j$, $x_j'$, $y_j$ and $y_j'$, we obtain 
\begin{eqnarray} 
{\cal N} & = & \left( \prod_{j = 1}^{l} \int d {\bf r}_j \, d {\bf 
r}_j' \, \tilde{F}_j ({\bf r}_j) \tilde{F}_j^* ({\bf r}_j') \right) 
\left( \prod_{j = 1}^{l'} \int d {\bf s}_j \, d {\bf s}_j' 
\tilde{G}_j^* ({\bf s}_j) \tilde{G}_j ({\bf s}_j') \right) 
\nonumber \\ 
&& \times \sum_{i, \, j, \, i', \, j'} \left( \prod_{j = 1}^{i} 
\sum_{{\bf p}_j} \frac{1}{\sqrt{2 E_{{p}_j} V}} \right) \left( 
\prod_{j = 1}^{i'} \sum_{{\bf q}_j} \frac{1}{\sqrt{2 E_{{q}_j} V}} 
\right) \left( \prod_{l = 1}^{j} \sum_{{\bf p}_j'} \frac{1}{\sqrt{2 
E_{{p}_j'} V}} \right) \left( \prod_{l = 1}^{j'} \sum_{{\bf q}_j'} 
\frac{1}{\sqrt{2 E_{{q}_j'} V}} \right) \nonumber \\ 
&& \times 2 \pi \delta [\sum^l r_{j 0} - \sum^{l'} s_{j 0} + \sum^i 
p_0 - \sum^{i'} q_0 ] \, 2 \pi \delta [\sum^{l'} s_{j 0}' - \sum^l 
r_{j 0}' + \sum^{j'} q_0' - \sum^j p_0'] \nonumber \\ 
&& \times V \delta (\sum^l {\bf r}_j - \sum^{l'} {\bf s}_j; 
\sum^{i'} {\bf q} - \sum^i {\bf p}) \, V \delta (\sum^{l'} {\bf 
s}_j' - \sum^l {\bf r}_j'; \, \sum^j {\bf p}' - \sum^{j'} {\bf q}') 
\nonumber \\ 
&& \times {\cal S} \, {\cal W} (\{ {\bf r}' \}, \{ {\bf p}' \}; \{ 
{\bf q}' \}, \{ {\bf s}' \} : \{ {\bf s} \}, \{ {\bf q} \}; \{ {\bf 
p} \}, \{ {\bf r} \}) e^{i [\sum^l (R_j - R_j') - \sum^{l'} (S_j - 
S_j') ] \cdot X_c} \, {\bf S} \, . 
\label{yatto3} 
\end{eqnarray} 
\narrowtext 

\noindent 
Note that, when $\langle S \rangle$ $(\in W)$ or $\langle S^\dagger 
\rangle$ consists of several disconnected parts, corresponding 
(momentum-conservation) $\delta$-function above becomes product of 
several $\delta$-functions. 

The form for ${\cal D}$, Eq.~(\ref{cal-D}), is given by 
Eq.~(\ref{yatto}) or Eq.~(\ref{yatto3}), in which factors related 
to the $\Phi$ fields are deleted. 

In general, ${\cal N}$ consists of several graphically disconnected 
parts. As assumed in Sec.~IIA, all $\Phi$'s are included in one 
connected parts ${\cal N}_{\mbox{\scriptsize{con}}}$. Other parts, 
which we write ${\cal D}$, include only the constituent particles 
$\phi$' of the system. Then, it is obvious that ${\cal N}$ takes the 
form ${\cal N} = {\cal N}_{\mbox{\scriptsize{con}}} {\cal D}$ (cf. 
Eq.~(\ref{hahaha})). It is also obvious that ${\cal D}$ is a 
contribution to ${\cal D}$ in Eqs.~(\ref{maru}). Then, such 
contribution does contribute to the reaction-probability ${\cal P}$, 
Eq.~(\ref{P}), as ${\cal N}_{\mbox{\scriptsize{con}}}$, which has 
already been dealt with in a lower-order level. Thus, computation of 
${\cal N}$'s, which consist of one connected part, is sufficient. 
\setcounter{equation}{0}
\setcounter{section}{2}
\section{Out-of-equilibrium reaction-probability formula} 
\def\theequation{\mbox{\arabic{section}.\arabic{equation}}}
\subsection{Preliminaries} 
In this section, we restrict our concern to quasiuniform systems 
near equilibrium and nonequilibrium quasistationary systems, which 
we simply refer to as out-of-equilibrium systems. Such systems 
are characterized \cite{chou} by weak dependence of the reaction 
probabilities on $X_c$ (cf. above after Eq.~(\ref{hadou})). More 
precisely, there exists a spacetime scale $L^\mu$, such that the 
reaction probabilities do not appreciably depend on $X_c$, when 
$X_c$ is in the spacetime region $|X_c^\mu - X_{c 0}^\mu|$ $(= 
|\Delta X_c^\mu|)$ $\lesssim L^\mu$ with $X_{c 0}^\mu$ an arbitrary 
spacetime point. For such systems, the reactions are regarded as 
taking place in the region $|X_c^\mu - X_{c 0}^\mu| \lesssim L^\mu$. 
Going to the momentum space, this means that the contribution (to 
the reaction probability ${\cal N}$) from the state that includes 
\lq\lq very soft'' momentum $|P^\mu| \lesssim 1 / L^\mu$ should be 
small. More precisely, the contribution from the summation-region in 
Eq.~(\ref{yatto}), in which at least one momentum (out of $\{ {\bf 
p}_j, {\bf q}_j,{\bf p}_j', {\bf q}_j' \}$) is \lq\lq very soft'' is 
negligibly small.\footnote{This is the case for most practical 
cases, which can be seen as follows. Let ${\cal T}$ be a typical 
scale(s) of the system under consideration. In the case of 
thermal-equilibrium system, ${\cal T}$ is the temperature of the 
system. Due to interactions, an effective mass is induced and the 
vacuum-theory mass $m$ turns out to the effective mass $M_{eff} 
(X_c)$. In the case of $m >> \sqrt{\lambda} {\cal T}$, $M_{eff} 
(X_c)$ is not much different from $m$ and, for $m \lesssim 
\sqrt{\lambda} {\cal T}$, a tadpole diagram induces mass of $O 
(\sqrt{\lambda} {\cal T})$, so that $M_{eff} (X_c) = O 
(\sqrt{\lambda} {\cal T})$. [$\sqrt{\lambda} {\cal T}$ (or even 
$\lambda {\cal T}$) is the scale that characterizes reactions. We 
assume that this scale is much larger than the \lq\lq very soft'' 
momentum scale, $1 / L^\mu << \sqrt{\lambda} {\cal T}$ (or $\lambda 
{\cal T}$).] Most amplitudes, when computed in perturbation theory 
(to be deduced below), are insensitive to the region $|P^\mu| \leq 
O (\sqrt{\lambda} {\cal T})$. Then, the contribution from the region 
$|P^\mu| \lesssim 1 / L^\mu$ is small, since the phase-space volume 
is small. Incidentally, in the case of equilibrium thermal QED or 
QCD $(m = 0)$, there are some quantities that diverge at infrared 
limits to leading order in hard-thermal-loop resummation scheme 
\cite{pis,le-b}. For such cases, more elaborate analysis is 
required.} 

Let us pick out $\langle a_{\bf p} \rangle $ from ${\cal S}$ in 
(\ref{haya}), which appears in ${\cal N}$, Eq.~(\ref{yatto}), in the 
form 
\begin{equation} 
\sum_{\bf p} \frac{1}{\sqrt{2 E_p V}} \, \langle a_{\bf p} \rangle 
\, e^{- i P \cdot \omega} \, , 
\label{oz} 
\end{equation} 
where $\omega$ stands for $\xi_j$ or $\zeta_j'$. The above 
observation shows that the quantity (\ref{oz}) does not appreciably 
depend on $\omega^\mu$, when $|\omega^\mu - X_{c 0}^\mu| \lesssim 
L^\mu$. This means that $\langle a_{\bf p} \rangle \simeq 0$ for 
$|p^i| \gtrsim 1 / L^i$ and $p^0 = E_p \gtrsim 1 / L^0$. Then, the 
argument at the end of the above paragraph shows that the 
contribution to ${\cal N}$ that include $\langle a \rangle$ can be 
ignored. Same reasoning shows that the contribution including 
$\langle a^\dagger \rangle$ and/or ${\cal S}_c (a a \, \cdots \, a 
)$ and/or ${\cal S}_c (a^\dagger a^\dagger \, \cdots \, a^\dagger)$ 
may also be ignored. 

Recalling the identity (\ref{ka}), we pick out from Eq.~(\ref{haya}) 
one ${\cal S}_c (a^\dagger_{{\bf p}_1} \, \cdots \, a^\dagger_{{\bf 
p}_j} a_{{\bf p}_{j + 1}} \, \cdots \, a_{{\bf p}_n} )$ ($n \geq 
3$). In ${\cal N}$ in Eq.~(\ref{yatto}), this factor appears in the 
form 
\begin{eqnarray} 
&& \sum_{\{ {\bf p} \}} \left( \prod_{l = 1}^n \frac{1}{\sqrt{2 
E_{p_l} V}} \right) {\cal S}_c \left( \left( \prod_{l = 1}^j 
a^\dagger_{{\bf p}_l} \right) \left( \prod_{l' = {j + 1}}^n 
a_{{\bf p}_{l'}} \right) \right) \nonumber \\ 
&& \mbox{\hspace*{3ex}} \times \exp \left[ i \left( 
\sum_{l = 1}^j P_l \cdot z_l - \sum_{l' = {j + 1}}^n P_{l'} 
\cdot z_{l'} \right) \right]  \, , 
\label{ic} 
\end{eqnarray} 
where $p_{l 0} = E_p$ $(l = 1, \cdots, n)$. It is not difficult to 
show that among the contributions to ${\cal N}$, there are 
contributions, whose counterparts of Eq.~(\ref{ic}), together with 
Eq.~(\ref{ic}), can be united into the form 
\begin{equation} 
{\cal C} (\{ z \}) \equiv i^{ n - 1} {\cal S}_c \left( : \phi (z_1) 
\cdots \phi (z_n) : \right) \, . 
\label{ic1} 
\end{equation} 
Here 
\[ 
\phi (z) = \sum_{\bf p} \frac{1}{\sqrt{2 E_p V}} \left[ a_{\bf 
p} e^{- i P \cdot z} + a_{\bf p}^\dagger e^{ i P \cdot z} \right] 
\, , 
\] 
where $p_0 = E_p$ and \lq $: \, \cdots \, :$' in Eq.~(\ref{ic1}) 
indicates to take the normal ordering. As discussed at the beginning 
of this subsection, for the system under consideration, the function 
(\ref{ic}) does not change appreciably in the region $|\Delta Z^\mu| 
\lesssim L^\mu$ ($Z = \sum_{l = 1}^n z_l / n$). This leads to an 
approximate momentum conservation for the function (\ref{ic}): 
\begin{equation} 
\Bigm| \sum_{l = 1}^j P_l^\mu - \sum_{l' = {j + 1}}^n P_{l'}^\mu 
\Bigm| \lesssim 1 / L^\mu \, . 
\label{mom} 
\end{equation} 
This is also the case for ${\cal C} (\{ z \})$ in Eq.~(\ref{ic1}). 
The conditions under which the initial correlations may 
be ignored are discussed in \cite{niea}. In the following, we ignore 
the initial correlations, inclusion of which into the formula 
obtained below is straightforward. 

After all this, in ${\cal S}$ in Eq.~(\ref{yatto}), we keep only 
$\langle a^\dagger a \rangle$'s: 
\widetext 
\begin{eqnarray} 
{\cal S} & = & \sum_{m, \, n} \sum_{gr} \langle a^\dagger_{{\bf 
p}_{l_j'}'} a_{{\bf q}_{k_{j'}'}'} \rangle \cdot \cdot \cdot 
\langle a^\dagger_{{\bf p}_{l_{n + 1}'}'} a_{{\bf q}_{k_{m + 1}'}'} 
\rangle \nonumber \\ 
&& \times \left( \delta_{{\bf q}_{k_{i'}}, \, {\bf q}_{k_m'}'} 
+ \langle a^\dagger_{{\bf q}_{k_{i'}}} a_{{\bf q}_{k_m'}'} \rangle 
\right) \cdot \cdot \cdot \left( \delta_{{\bf q}_{k_{i - n + 1}}, \, 
{\bf q}_{k_1'}'} + \langle a^\dagger_{{\bf q}_{k_{i - n + 1}}} 
a_{{\bf q}_{k_1'}'} \rangle \right) \nonumber \\ 
&& \times \langle a^\dagger_{{\bf q}_{j_{i - n}}} 
a_{{\bf p}_{l_i}} \rangle \cdot \cdot \cdot \langle a^\dagger_{{\bf 
q}_{j_1}} a_{{\bf p}_{l_{n + 1}}} \rangle \, \langle a^\dagger_{{\bf 
p}_{l_n'}'} a_{{\bf p}_{l_n}} \rangle \cdot \cdot \cdot \langle 
a^\dagger_{{\bf p}_{l_1'}'} a_{{\bf p}_{l_1}} \rangle \, , 
\label{SS} 
\end{eqnarray} 
\narrowtext 

\noindent 
where $j - n = j' - m$ and $i' - m = i - n$, which leads to $i + j' 
= j + i'$. 

Referring to (\ref{yatto}), we use the following set-symbols 
throughout in the sequel: 
\begin{eqnarray*} 
&& {\cal V}_\Phi =  {\cal V}_\Phi^S \cup {\cal V}_\Phi^{S^\dagger} 
\, ; 
\;\;\; {\cal V}_\Phi^S = \{ x \} \cup \{ y \} \, , \;\; 
{\cal V}_\Phi^{S^\dagger} = \{ x' \} \cup \{ y' \} \, , \nonumber 
\\ 
&& {\cal V}_e = {\cal V}_e^S \cup {\cal V}_e^{S^\dagger} \, ; 
\;\;\; {\cal V}_e^S = \{ \xi \} \cup \{ \zeta \} \, , \;\; 
{\cal V}_e^{S^\dagger} = \{ \xi' \} \cup \{ \zeta' \} \, ,  
\end{eqnarray*} 
and ${\cal V}_i = {\cal V}_i^S \cup {\cal V}_i^{S^\dagger}$ with 
${\cal V}_i^S$ [${\cal V}_i^{S^\dagger}$] the set of internal-vertex 
points in $\langle S \rangle$ [$\langle S^\dagger \rangle$] ($\in 
{\cal W}$). When the vertex point $\xi_j$ ($\xi_j'$) or $\zeta_l$ 
($\zeta_l'$) coincides with one of the vertex points in ${\cal 
V}_\Phi^S$ (${\cal V}_\Phi^{S^\dagger}$), we include it in ${\cal 
V}_\Phi^S$ (${\cal V}_\Phi^{S^\dagger}$). At the final stage, ${\cal 
V}_\Phi$ (${\cal V}_e \cup {\cal V}_i$) turns out to the set of 
external-vertex (internal-vertex) points of the out-of-equilibrium 
amplitude (\ref{owayo}) representing ${\cal P}$. 
\subsection{Two-point function} 
>From Eq.~(\ref{yatto}) with Eq.~(\ref{SS}), we pick out 
\[ 
i \tilde{\Delta} (\rho, \sigma) \equiv \sum_{{\bf p}, \, {\bf p}'} 
\frac{1}{\sqrt{2 E_p V} \sqrt{2 E_{p'} V}} \langle a^\dagger_{{\bf 
p}'} \, a_{\bf p} \rangle \, e^{- i (P \cdot \rho - P' 
\cdot \sigma)} \, , 
\] 
where $p_0 = E_p$, $p_0' = E_{p'}$, and ${\bf p} \in \{ {\bf p} \} 
\cup \{ {\bf q}' \}$, ${\bf p}' \in \{ {\bf p}' \} \cup \{ {\bf q} 
\}$, $\rho \in \{ \xi \} \cup \{ \zeta' \}$ and $\sigma \in \{ \xi' 
\} \cup \{ \zeta \}$. Changing ${\bf p}$ and ${\bf p}'$ to 
\[ 
{\bf p}_+ = ({\bf p} + {\bf p}' ) / 2 \, , \;\;\;\; \; 
{\bf p}_- = {\bf p} - {\bf p}' \, , \;\;\;\; \; 
\] 
we get 
\begin{eqnarray} 
i \tilde{\Delta} (\rho, \sigma) & = & \sum_{{\bf p}_+} 
\frac{1}{\sqrt{2 E_+ V} \sqrt{2 E_- V}} \, e^{- i P_+ \cdot (\rho - 
\sigma)} \nonumber \\ 
&& \times \tilde{N} \left(X, {\bf p}_+ \right) \, , 
\label{2-2} 
\\ 
\tilde{N} \left( X; {\bf p}_+ \right) & = & \sum_{{\bf p}_-} \, e^{ 
- i (E_+ - E_-) X_0} e^{i {\bf p}_- \cdot {\bf X}} \nonumber \\ 
&& \times \langle a^\dagger_{{\bf p}_+ - {\bf p}_- / 2} \, a_{{\bf 
p}_+ + {\bf p}_- / 2} \rangle \, , 
\label{2-3}
\end{eqnarray} 
where $X = (\rho + \sigma) / 2$, $E_{\pm} = E_{|{\bf p}_+ \mp i 
\nabla_{\bf X} / 2|}$ and $p_+^0 = (E_+ + E_-) / 2$. It is worth 
mentioning in passing that one can easily derive from 
Eq.~(\ref{2-3}) $P \cdot \partial_X \tilde{N} = 0$. 

Now, Eq.~(\ref{2-2}) may be written as 
\begin{eqnarray} 
i \tilde{\Delta} (\rho, \sigma) & = & 
\sum {\mbox{\hspace*{-0.525cm}}} \int {\cal D}^{\, 4} P \, 
e^{ -i P \cdot (\rho - \sigma)} \, \frac{p_0}{\sqrt{E_+ E_-}} 
\, 2 \pi \theta (p_0) \nonumber \\ 
&& \times \delta \left( p^2_0 - \left( \frac{E_+ + E_-}{2} \right)^2 
\right) \, \tilde{N} (X ; {\bf p}) \, , 
\nonumber \\ 
\label{waru} 
\end{eqnarray} 
where 
\[ 
\sum {\mbox{\hspace*{-0.525cm}}} \int {\cal D}^{\, 4} P \equiv \int 
\frac{d p_0}{2 \pi} \sum_{\bf p} \frac{1}{V} \, . 
\] 
As usual, we rewrite $p = |{\bf p}|$ in terms of $p_0$ by using 
$\delta_+ (p_0^2 - \cdots \, )$ in Eq.~(\ref{waru}). In doing so we 
obtain 
\[ 
\frac{p_0}{\sqrt{E_+ E_-}} \tilde{N} (X ; {\bf p}) \rightarrow 
\left[ 1 + \frac{({\bf p} \cdot \nabla_{\bf X})^2}{4 p_0^4} 
\right]^{- 1 / 2} \tilde{N} (X; p_0, \hat{\bf p}) \, , 
\] 
where $\hat{\bf p} \equiv {\bf p} / p$. Carrying out the derivative 
expansion (expansion with respect to $\partial_{X_\mu}$) and keeping 
up to the second-order ${\bf X}$-derivative terms, we obtain 
\begin{eqnarray} 
i \tilde{\Delta} (\rho, \sigma) & = & 
\sum {\mbox{\hspace*{-0.525cm}}} \int {\cal D}^{\, 4} P \, 
e^{ -i P \cdot (\rho - \sigma)} \nonumber \\ 
&& \times 2 \pi \delta_+ (P^2 - m^2) \, N (X ; p_0, \hat{\bf p}) 
\, , \label{4-8} 
\end{eqnarray} 
where $\delta_+ (P^2 - m^2) = \theta (p_0) \delta (P^2 - m^2)$ and 
\begin{eqnarray} 
N (X ; p_0, \hat{\bf p}) & = & \left[ 1 - \frac{1}{4 } 
\frac{\stackrel{\leftarrow}{\partial}}{\partial m^2} \left( 
\nabla_{\bf X}^2 - ( {\bf v} \cdot \nabla_{\bf X})^2\right) 
\right. \nonumber \\ 
&& \left. - \frac{({\bf v} \cdot \nabla_{\bf X})^2}{8 p_0^2} + 
\cdots \, \right] \, \tilde{N} (X; p_0, \hat{\bf p}) \, . 
\label{B} 
\end{eqnarray} 
Here $\stackrel{\leftarrow}{\partial} / \partial m^2$ acts on 
$\delta_+ (P^2 - m^2)$ in Eq.~(\ref{4-8}) and ${\bf v} = {\bf p} / 
p_0$. 

In case of the system in which translation invariance holds, 
$\langle a^\dagger_{\bf p} a_{\bf q} \rangle_c \propto 
\delta_{{\bf p}, \,{\bf q}}$, Eq.~(\ref{2-3}) tells us that $N (p_0, 
\hat{\bf p}) = \tilde{N} (p_0, \hat{\bf p})$ is the number 
density of a particle with momentum ${\bf p}$. This allows us to 
interpret $N (X; p_0, \hat{\bf p})$ as the \lq\lq bare'' number 
density of a quasiparticle with ${\bf p}$ at the spacetime point 
$X^\mu$. (For more details, see \cite{niea}.) 
\subsection{Construction of out-of-equilibrium propagators} 
So far, for simplicity of presentation, we have dealt with 
real-scalar-field systems. Physical meaning of the propagators to be 
deduced below can be determined more transparent manner by employing 
a complex-scalar-field systems, which we deal with in the sequel of 
this section. Let $a_{\bf p}$ ($a_{\bf 
p}^\dagger$) be an annihilation [a creation] operator for a particle 
of momentum ${\bf p}$. The antiparticle counterpart of $a_{\bf p}$ 
($a_{\bf p}^\dagger$) is $b_{\bf p}$ ($b_{\bf p}^\dagger$). For 
simplicity, we assume that the density-matrix operator $\rho$ 
commutes with 
charge operator $Q$, $[\rho, Q] = 0$. Then, all but $\langle a_{\bf 
p}^\dagger a_{\bf q} \rangle$, $\langle b_{\bf p}^\dagger b_{\bf q} 
\rangle$, $\langle a_{\bf p}^\dagger b_{\bf q}^\dagger \rangle$, 
$\langle a_{\bf p} b_{\bf q} \rangle$ vanish. Same reasoning as at 
the beginning of this section shows that $\langle a_{\bf p}^\dagger 
b_{\bf q}^\dagger \rangle$ and $\langle a_{\bf p} b_{\bf q} \rangle$ 
are negligibly small. Thus, we are left with $\langle a_{\bf 
p}^\dagger a_{\bf q} \rangle$'s and $\langle b_{\bf p}^\dagger b_{\bf 
q} \rangle$'s. 

A) Let us take a Feynman diagram ${\cal F}$ for ${\cal N}$ (cf. 
Eq.~(\ref{yatto})), and pick out from ${\cal F}$ a vacuum-theory 
propagator $i \Delta^{(0)} (z_1 - z_2) = \langle 0 | \, T \phi (z_1) 
\phi^\dagger (z_2)\, | 0 \rangle$ $\in \langle S \rangle (\in {\cal 
N})$. Then, we pick up the following two diagrams for ${\cal N}$. 
The first one is the same as ${\cal F}$, except that $i \Delta^{(0)} 
(z_1 - z_2)$ is replaced by 
\[ 
\sum_{\bf p} \frac{1}{\sqrt{2 E_{p} V}} e^{- i P \cdot z_1} 
\sum_{\bf q} \frac{1}{\sqrt{2 E_{q} V}} e^{i Q \cdot z_2} \langle 
a^\dagger_{\bf q} a_{\bf p} \rangle \, , 
\] 
which is involved in Eq.~(\ref{yatto}). The second one is the same 
as ${\cal F}$, except that $i \Delta^{(0)} (z_1 - z_2)$ is replaced 
by 
\[ 
\sum_{\bf q} \frac{1}{\sqrt{2 E_{q} V}} e^{- i \overline{Q} \cdot 
z_2} \sum_{\bf p} \frac{1}{\sqrt{2 E_{p} V}} e^{i \overline{P} \cdot 
z_1} \langle b^\dagger_{- {\bf p}} b_{- {\bf q}} \rangle \, , 
\] 
with $\overline{P} \equiv (E_p, - {\bf p})$, etc. Adding the above 
two contributions to the original contribution, and Fourier 
transforming on $z_1 - z_2$, we obtain for the relevant part, 
\begin{eqnarray} 
&& i \Delta_{1 1} \left( \frac{z_1 + z_2}{2}; P \right) \nonumber \\ 
&& \mbox{\hspace*{4ex}} \equiv \frac{i}{P^2 - m^2 + i 0^+} 
\nonumber \\ 
&& \mbox{\hspace*{7ex}} + 2 \pi \delta (P^2 - m^2) N 
\left( \frac{z_1 + z_2}{2}; p_0, \hat{\bf p} \right) \, . 
\label{e11} 
\end{eqnarray} 
Here, $N$ with $p_0 > 0$ is as in Eq.~(\ref{B}) with (\ref{2-3}), 
while, for $p_0 < 0$, $N$ takes the same form (\ref{B}) where 
$\tilde{N}$ is defined, with obvious notation, as 
\begin{eqnarray*} 
&& \tilde{N} \left( \frac{z_1 +z_2}{2}; p_0, \hat{\bf p} \right) 
\nonumber \\ 
&& \mbox{\hspace*{4ex}} = \sum_{{\bf p}_-} \, e^{i (E_+ - E_-) 
(z_{1 0} + z_{2 0}) / 2} \nonumber \\ 
&& \mbox{\hspace*{7ex}} \times e^{- i {\bf p}_- \cdot ({\bf z}_1 + 
{\bf z}_2) / 2} \, \langle b^\dagger_{ - {\bf p} + {\bf p}_- / 2} \, 
b_{- {\bf p} - {\bf p}_- / 2} \rangle 
\end{eqnarray*} 
with, as before, $E_{\pm} = E_{|{\bf p}_+ \mp i \nabla_{\bf X} / 
2|}$. 

As discussed at the end of the last subsection, $N (X; p_0, \hat{\bf 
p})$ with $p_0 > 0$ is the \lq\lq bare'' number density of a 
quasiparticle with momentum ${\bf p}$ at the point $X^\mu$. 
Similarly, $N (X; p_0, \hat{\bf p})$ with $p_0 < 0$ is the \lq\lq 
bare'' number density of an anti-quasiparticle with momentum $- {\bf 
p}$ at $X^\mu$. 

B) Starting from $\langle S^\dagger \rangle$ $(\in {\cal N})$ that 
includes a vacuum-theory propagator $[i \Delta^{(0)} (z_1 - z_2)]^*$ 
and proceeding as above A), we obtain 
\begin{eqnarray} 
i \Delta_{2 2} (X; P) & \equiv & [i \Delta_{1 1} (X; P)]^* \nonumber 
\\ 
& = & \frac{- i}{P^2 - m^2 - i 0^+} \nonumber \\ 
&& + 2 \pi \delta (P^2 - m^2) 
N \left( \frac{z_1 + z_2}{2}; p_0, \hat{\bf p} \right) \, . 
\nonumber \\ 
\label{e22} 
\end{eqnarray} 

C) Let us take a set of Feynman diagrams ${\cal F}_1$ and ${\cal 
F}_2$. ${\cal F}_1$ contains (cf. Eq.~(\ref{SS})) 
\begin{eqnarray} 
&& \sum_{{\bf p}_j} \frac{1}{\sqrt{2 E_{p_j} V}} e^{- i P_j \cdot 
z_1} \sum_{{\bf q}_k} \frac{1}{\sqrt{2 E_{q_k} V}} 
e^{i Q_k \cdot z_2} \nonumber \\ 
&& \mbox{\hspace*{4ex}} \times \left( \delta_{{\bf q}_k, \, {\bf 
p}_j} + \langle a^\dagger_{{\bf q}_k} a_{{\bf p}_j} \rangle \right) 
\label{irekae2} 
\end{eqnarray} 
with $z_1 \in \{ \zeta' \}$ $(\in \langle S^\dagger \rangle)$ and 
$z_2 \in \{ \zeta \}$ $(\in \langle S \rangle)$. ${\cal F}_2$ is 
same as ${\cal F}_1$ except that (\ref{irekae2}) is replaced by 
\[ 
\sum_{{\bf q}_k} \frac{1}{\sqrt{2 E_{q_k} V}} e^{- i 
\overline{Q}_k \cdot z_2} \sum_{p_j} \frac{1}{\sqrt{2 E_{p_j} V}} 
e^{i \overline{P}_j \cdot z_1} \langle b^\dagger_{- {\bf p}_j} b_{- 
{\bf q}_k} \rangle 
\] 
with $z_1 \in \{ \xi' \}$ $(\in \langle S^\dagger \rangle)$ and $z_2 
\in \{ \xi \}$ $(\in \langle S \rangle)$. Adding the contributions 
from ${\cal F}_1$ and from ${\cal F}_2$, we extract the relevant 
part, of which the Fourier transformation on $z_1 - z_2$ is 
\begin{eqnarray} 
i \Delta_{2 1} (X; P) & \equiv & 2 \pi \delta (P^2 - m^2) \nonumber 
\\ 
&& \times \left[ \theta (p_0) + N \left( \frac{z_1 + z_2}{2}; p_0, 
\hat{\bf p} \right) \right] \, . 
\label{e21} 
\end{eqnarray} 

D) Let us take a set of Feynman diagrams ${\cal F}_1'$ and ${\cal 
F}_2'$. ${\cal F}_1'$ contains 
\begin{eqnarray} 
&& \sum_{{\bf p}_j} \frac{1}{\sqrt{2 E_{p_j} V}} e^{- i P_j \cdot 
z_1} \sum_{{\bf q}_k} \frac{1}{\sqrt{2 E_{q_k} V}} e^{i Q_k \cdot 
z_2} \langle a^\dagger_{{\bf q}_k} a_{{\bf p}_j} \rangle \nonumber 
\\ 
\label{irekae6} 
\end{eqnarray} 
with $z_1 \in \{ \xi \}$ $(\in \langle S \rangle)$ and $z_2 \in \{ 
\xi' \}$ $(\in \langle S^\dagger \rangle)$. ${\cal F}_2'$ is same 
as ${\cal F}_1'$ except that (\ref{irekae6}) is replaced by 
\begin{eqnarray*} 
&& \sum_{{\bf q}_k} \frac{1}{\sqrt{2 E_{q_k} V}} e^{- i 
\overline{Q}_k \cdot z_2} \sum_{{\bf p}_j} \frac{1}{\sqrt{2 E_{p_j} 
V}} e^{i \overline{P}_j \cdot z_1} \nonumber \\ 
&& \mbox{\hspace*{4ex}} \times \left( \delta_{{\bf p}_j, \, {\bf 
q}_k} + \langle b^\dagger_{- {\bf p}_j} b_{- {\bf q}_k} \rangle 
\right) 
\end{eqnarray*} 
with $z_1 \in \{ \zeta \}$ $(\in \langle S \rangle)$ and $z_2 \in \{ 
\zeta' \}$ $(\in \langle S^\dagger \rangle)$. Adding the 
contributions from ${\cal F}_1'$ and from ${\cal F}_2'$, we extract 
the relevant part, of which the Fourier transformation on $z_1 - 
z_2$ is 
\begin{eqnarray} 
i \Delta_{1 2} (X; P) & \equiv & 2 \pi \delta (P^2 - m^2) \nonumber 
\\ 
&& \times \left[ \theta 
(- p_0) + N \left( \frac{z_1 + z_2}{2}; p_0, \hat{\bf p} 
\right) \right] \, . 
\label{e12} 
\end{eqnarray} 

Above derivation of $i \Delta_{i j}$ $(i, \, j = 1, 2)$ is self 
explanatory for their physical meaning or interpretation. The 
physical interpretation is summarized as generalized cutting rules, 
which is a generalization of Cutkosky's cutting rules in vacuum 
theory. (For more details, see \cite{niesin}.) 
\subsection{Closed-time-path formalism} 
$i \Delta_{i j}$ $(i, \, j = 1, 2)$ obtained above are nothing but 
the propagators in the closed-time-path (CTP) formalism of 
out-of-equilibrium quantum field theory \cite{chou}. The CTP 
formalism is constructed on the directed time-path $C = C_1 \oplus 
C_2$ in a complex-time plane, where $C_1 = (- \infty \to + \infty)$ 
and $C_2 = (+ \infty \to - \infty)$. A field $\phi (x_0, {\bf x})$ 
with $x_0 \in C_1$  [$x_0 \in C_2$] is denoted by $\phi_1 (x_0, {\bf 
x})$ [$\phi_2 (x_0, {\bf x})$] and is called a type-1 [type-2] 
field. The interaction Lagrangian density is of the form, 
\begin{eqnarray*} 
{\cal L}_{\mbox{\scriptsize{int}}} & = & {\cal 
L}_{\mbox{\scriptsize{int}}}^{(1)} - {\cal 
L}_{\mbox{\scriptsize{int}}}^{(2)} \, , \nonumber \\ 
{\cal L}_{\mbox{\scriptsize{int}}}^{(i)} & = & - \frac{\lambda}{4} 
(\phi^\dagger_i \phi_i)^2 - \frac{g}{(n!)^2} \Phi_i (\phi^\dagger_i 
\phi_i)^n \, , \;\;\;\; (i = 1, 2) \, . 
\end{eqnarray*} 
Then, the vertex factor for the \lq\lq type-1 vertex'' that comes 
from ${\cal L}_{int}^{(1)}$ is the same as in vacuum theory, while 
the vertex factor for the \lq\lq type-2 vertex'' is minus the 
corresponding \lq\lq type-1 vertex factor.'' The CTP propagators are 
defined by the statistical average of the time-path-ordered product 
of fields, which are written as 
\begin{eqnarray} 
i \Delta_{1 1} (x, y) & = & \langle \mbox{T} \phi_1 (x) 
\phi^\dagger_1 (y) \rangle_c \, , \nonumber \\ 
i \Delta_{2 2} (x, y) & = & \langle \overline{\mbox{T}} \phi_2 (x) 
\phi^\dagger_2 (y) \rangle_c = [i \Delta_{1 1} (y, x)]^* \, , 
\nonumber \\ 
i \Delta_{1 2} (x, y) & = & \langle \phi_2^\dagger (y) 
\phi_1 (x) \rangle_c \, , \nonumber \\ 
i \Delta_{2 1} (x, y) & = & \langle \phi_2 (x) 
\phi^\dagger_1 (y) \rangle_c \, , 
\label{tata} 
\end{eqnarray} 
where T ($\overline{\mbox{T}}$) is the time-ordering 
(anti-time-ordering) symbol. In computing (\ref{tata}), one 
identifies $\phi_2$ with $\phi_1$. Comparing Eq.~(\ref{tata}) with 
the above deduction of $\Delta_{i j}$ $(i, j = 1, 2)$, 
Eqs.~(\ref{e11}), (\ref{e22}), (\ref{e21}), and (\ref{e12}), we see 
that $x$ of $\phi_1 (x)$ in Eq.~(\ref{tata}) corresponds to a 
vertex-point in $\langle S \rangle$ $(\in {\cal W})$ and $x$ of 
$\phi_2 (x)$ corresponds to a vertex-point in $\langle S^\dagger 
\rangle$. The vertex factors in $\langle S \rangle$ ($\in {\cal W}$) 
are $- i \lambda$ for $- \lambda (\phi^\dagger \phi)^2 / 4$ 
interaction and $- i g$ for $- g \Phi (\phi^\dagger \phi)^n / 
(n!)^2$ interaction. Then, the vertex factors in $\langle S^\dagger 
\rangle$ $(\in {\cal W})$ are, in corresponding order to the above, 
$i \lambda$ and $i g$. This is in accord with the above-mentioned 
vertex factors in the CTP formalism. 
\subsection{Reaction-probability formula} 
Observation made so far shows that ${\cal N}$ in Eq.~(\ref{yatto}) 
with Eq.~(\ref{SS}) corresponds to an amplitude in the CTP formalism 
of the \lq\lq process,'' 
\begin{equation} 
\sum_{j = 1}^l \Phi_{1 j} + \sum_{j = 1}^{l'} \Phi_{2 j} \to 
\sum_{j = 1}^l \Phi_{2 j} + \sum_{j = 1}^{l'} \Phi_{1 j} \, . 
\label{AA} 
\end{equation} 
As mentioned at the end of Sec.~II, only connected ${\cal N}$'s 
contribute to the reaction-probability ${\cal P}$. Thus, we finally 
obtain 
\begin{eqnarray} 
{\cal P} & = & \left( \prod_{j = 1}^{l} \int d^{\, 4} x_j \, d^{\, 
4} x_j' F_j (x_j) F^* (x_j') \right) \nonumber \\ 
&& \times \left( \prod_{j = 1}^{l'} \int 
d^{\, 4} y_j \, d^{\, 4} y_j' G_j^* (y_j) G_j (y_j') \right) 
\nonumber \\ 
&& \times \sum_{\mbox{\scriptsize{diagrams}}} \int d^{\, 4} 
\omega_1 \, \cdots \, \omega_{N_d} \, {\cal F}_i (X; \{ (\omega_k - 
\omega_{k'}) \} ) \, , \nonumber \\ 
\label{owayo} 
\end{eqnarray} 
where ${\cal F}_i$ is a {\em connected} amplitude in the CTP 
formalism which includes all $\Phi$'s. In Eq.~(\ref{owayo}), we have 
used $\{ \omega \}$ for collectively denoting all the (external and 
internal) vertex-points and the summation runs over diagrams. A pair 
of $\omega$'s, $\omega_k$ and $\omega_{k'}$, in a pair of brackets 
$(\, \cdots \,)$ in ${\cal F}_i$ denotes the vertex-points that are 
connected by $i \Delta_{k l} (\omega_{k (k')}, \omega_{k' (k)})$. 

Here some remarks are in order. 
\begin{description} 
\item{1)} As mentioned at the beginning of section, inclusion of the 
initial correlations (\ref{ic}) or (\ref{ic1}) is straightforward. 
\item{2)} Taking the infinite-volume limit $V \to \infty$ goes as 
follows: 
\begin{eqnarray*} 
&& \sum_{\bf p} \rightarrow \frac{V}{(2 \pi)^3} \int d^{\, 3} p \, , 
\\ 
&& a_{\bf p} \rightarrow \sqrt{\frac{(2 \pi)^3}{V}} a ({\bf p})\, , 
\;\;\; \mbox{etc.} 
\end{eqnarray*} 
Above deduction shows that there is no finite-volume correction, in 
the sense that there do not exist extra contributions to ${\cal N}$, 
which disappear in the limit $V \to \infty$. It should be stressed 
that this statement holds for periodic boundary conditions. 
\item{3)} It is clear from the above deduction (cf. Subsecs. B and 
C) that the CTP formalism here is formulated in terms of the \lq\lq 
bare'' number density of quasiparticles. A canonical CTP formalism 
is formulated in terms of the physical or observed number density of 
quasiparticles. How to translate the former into the latter is 
discussed in \cite{niea}. 
\end{description} 

Finally, we make a comment on gauge theories. If we choose a 
physical gauge like the Coulomb gauge or the Landshoff-Rebhan 
variant \cite{lr} of a covariant gauge, the gauge boson may be dealt 
with in a similar manner to the above scalar-field case. If we adopt 
a traditional covariant gauge, a straightforward modification is 
necessary. 
\setcounter{equation}{0}
\setcounter{section}{3}
\section{Computational procedure} 
\def\theequation{\mbox{\arabic{section}.\arabic{equation}}}
In this section, we present a concrete procedure of computing the 
reaction probability ${\cal P}$ up to $n$th-order terms with 
respect to the $X_\mu$ derivatives. 

1) From ${\cal F}_i$ in Eq.~(\ref{owayo}), we pick out $i \Delta_{i 
j} (\rho, \sigma)$, 
\begin{eqnarray} 
\Delta_{i j} (\rho, \sigma) = 
\sum {\mbox{\hspace*{-0.525cm}}} \int {\cal D}^{\, 4} P \, 
e^{- i P \cdot (\rho - \sigma)} 
\Delta_{i j} \left( \frac{\rho + \sigma}{2}; P \right) \, . 
\label{shuppatsu} 
\end{eqnarray} 
Since ${\cal F}_i$ includes $\Phi$'s, the vertex-point $\rho$ 
[$\sigma$] is connected\footnote{Note that, in general, the 
vertex-points $v$ and $v'$ are not uniquely singled out. ($v$ can 
coincides with $v'$.) However, different choices of $v$ and $v'$ 
leads to the same reaction probability ${\cal P}$ within the 
accuracy under consideration.} with a vertex-point $v$ [$v'$] 
$\in{\cal V}_\Phi$ (cf. Fig.~1): 
\begin{eqnarray} 
\frac{\rho + \sigma}{2} & = & \frac{1}{2} \left[ - 
\sum_{j = 0}^k (\omega_{j + 1} - \omega_j) \right. \nonumber \\ 
&& \left. + \sum_{j = 0}^{k'} (\omega_j' - \omega_{j + 1}') + v + 
v' \right] \, , 
\label{chain1} 
\end{eqnarray} 
where $\omega_0 = \rho$, $\omega_0' = \sigma$, $\omega_{k + 1} = v$, 
$\omega_{k' + 1}' = v'$, with $v, \, v' \in {\cal V}_\Phi$. In 
Eq.~(\ref{chain1}), each pair of spacetime points in a pair of 
brackets, $\omega_{j + 1}$ and $\omega_j$ [$\omega_j'$ and 
$\omega_{j + 1}'$], is connected by one or several $i \Delta_{k l} 
(\omega_{j + 1}, \omega_j)$ [$i \Delta_{k' l'} (\omega_j', \omega_{j 
+ 1}')$] in ${\cal F}_i$ (cf. Fig.~1). Here, we note that $v$ and 
$v'$ may be written as 
\begin{equation}  
v = X + \tilde{v} \, , \;\;\;\;\;\; v' = X + \tilde{v}' \, , 
\label{mid} 
\end{equation}  
where $X$ is the mid-point of the external-vertex points, around 
which the reaction is taking place: 
\[ 
X = \frac{1}{2 (l + l')} \left[ \sum_{j = 1}^l (x_j + x_j') + 
\sum_{j = 1}^{l'} (y_j + y_j') \right] \, . 
\] 

2) Using Eqs.~(\ref{chain1}) and (\ref{mid}), we expand $\Delta_{i 
j} ((\rho + \sigma) / 2; P)$ in Eq.~(\ref{shuppatsu}) as 
\begin{eqnarray} 
&& \Delta_{i j} \left( \frac{\rho + \sigma}{2}; P \right) \nonumber 
\\ 
&& \mbox{\hspace*{3ex}} = \Delta_{i j} (X; P) + \frac{1}{2} \Big[ - 
\sum_{j = 0}^k (\omega_{j + 1} - \omega_j) \nonumber \\ 
&& \mbox{\hspace*{6ex}} + \sum_{j = 0}^{k'} (\omega_j' - 
\omega_{j + 1}') + \tilde{v} + \tilde{v}' \Big] \cdot \partial_X 
\Delta_{i j} 
(X; P) + \, \cdots \, , \nonumber \\ 
\label{ten} 
\end{eqnarray} 
where \lq $\, \cdots \,$' stands for terms with higher-order 
derivative with respect to $X$. The series (\ref{ten}) is truncated 
at the $n$th-order terms with respect to the $X_\mu$ derivatives. 
The approximation in which \lq $\, \cdots \,$ is ignored is called 
the gradient approximation. 

3) Let us deal with the term with $(\omega_{j + 1} - \omega_j)$ in 
Eq.~(\ref{ten}). It can easily be shown that $(\omega_{j + 1} - 
\omega_j) \, i \Delta_{k l} (\omega_{j + 1}, \omega_j)$ 
becomes\footnote{As in the case of some self-energy-type subdiagram, 
there are several $i \Delta_{k l} (\omega_{j + 1}, \omega_j)$'s [$i 
\Delta_{k' l'} (\omega_j', \omega_{j + 1}')$'s] (cf. Fig.~1). In 
such a case, one chooses any one of them.} 
\widetext 
\begin{eqnarray*} 
&& (\omega_{j + 1} - \omega_j)^\mu 
\sum {\mbox{\hspace*{-0.525cm}}} \int {\cal D}^{\, 4} P' \, 
e^{- i (P + P') \cdot (\omega_{j + 1} 
- \omega_j) } i \Delta_{k l} \left( \frac{\omega_{j + 1} + 
\omega_j}{2}; P + P' \right) \nonumber \\ 
& & \mbox{\hspace*{4ex}} = 
\sum {\mbox{\hspace*{-0.525cm}}} \int {\cal D}^{\, 4} P' \, 
e^{- i (P + P') \cdot (\omega_{j + 1} - 
\omega_j)} \frac{\partial}{i \partial P_\mu'} i \Delta_{k l} \left( 
\frac{\omega_j + \omega_{j + 1}}{2}; P + P' \right) \, . 
\end{eqnarray*} 
\narrowtext 

\noindent 
Other terms and higher $X^\mu$-derivative terms \lq $\, \cdots \, 
$' in Eq.~(\ref{ten}) may be dealt with similarly. All other parts 
of ${\cal F}_i$, Eq.~(\ref{owayo}), than the one (\ref{shuppatsu}) 
may be dealt with similarly. 

4) Carrying out the integrations over all vertex-points except those 
in ${\cal V}_\Phi$, we have momentum-conservation $\delta$-functions 
at each internal vertex point. 

As discussed at the beginning of Sec.~III, the wave functions of 
$\Phi$'s should be localized within the space region $\lesssim L^i$ 
$(i = 1, 2, 3)$. However, for simplicity, we assume in the sequel 
that the wave functions of $\Phi$'s are of plane-wave 
form,\footnote{It is to be noted that, if we use the the plane-wave 
form (\ref{pl}) in Eq.~(\ref{yatto3}), $X$-dependence disappears. In 
the procedure presented here, $X$-dependence of ${\cal N}$ is 
already (partially) taken into account before arriving at 4).} 
\begin{eqnarray} 
F_j (x) & = & e^{- i R_j \cdot x} / \left( 2 V \sqrt{r_j^2 + M^2} 
\right)^{1 / 2} \, , \nonumber \\ 
G_j (y) & = & e^{- i R_j' \cdot y} /  \left( 2 V \sqrt{r_j^{' \, 2} 
+ M^2} \right)^{1 / 2} \, . 
\label{pl} 
\end{eqnarray} 

5) We carry out the integrations over all vertex-points in ${\cal 
V}_\Phi$ to yield momentum-conservation $\delta$-functions at those 
vertex points and we are left with integrations over the independent 
or loop momenta. Keeping the terms up to the $n$th-order terms with 
respect to the $X_\mu$ derivatives, we obtain the final formula, 
which may be written in the form, 
\begin{equation} 
{\cal P} = \int d^{\, 4} X \, A (X; R_1', \, \cdots \, , R_{l'}'; 
R_1, \cdots , R_l) \, . 
\label{ti} 
\end{equation} 
Note that $A$ depends weakly on $X$ through $N (X; Q_k)$'s. From 
Eq.~(\ref{ti}), we see that $A$ is the reaction rate per unit 
volume. Incidentally, were it not for this $X$-dependence, 
integration over $X$ in Eq.~(\ref{ti}) would yield $V T$, where $V$ 
is the volume of the system and $T = t_f - t_i$ is the time interval 
during which the reaction takes place. In the limit $V, \, T \to 
\infty$, the $V T$ becomes 
\[ 
\lim_{V, \, T \to \infty} V \, T = (2 \pi)^4 \delta^{\, 4} (0) \, . 
\] 
\subsection*{Example} 
Here, for the purpose of illustration, we deal with the 
heavy-$\Phi$ production process, 
\begin{equation} 
\mbox{Out-of-equilibrium system} \to \Phi + \mbox{anything} \, . 
\label{reac} 
\end{equation} 
The system is composed of real scalar $\phi$ with ${\cal 
L}_{\mbox{\scriptsize{int}}} = - \lambda \phi^3 / 3!$, and $\Phi$ 
interacts with $\phi$ through ${\cal L}_{\phi \Phi} = - g \Phi 
\phi^2 / 2$. We analyze the contribution from Fig.~2 for ${\cal P}$ 
in Eqs.~(\ref{maru}). Using Eq.~(\ref{yatto}), we have 
\begin{eqnarray*} 
{\cal N} & = & g^2 \lambda^2 \int d^{\, 4} x' G^* (x') \int 
d^{\, 4} y' G (y') \\ 
& & \times \sum_{{\bf p}_1} \frac{1}{\sqrt{2 E_{p_1} V}} \, e^{- i 
P_1 \cdot x'} \int d^{\, 4} \xi \sum_{{\bf p}_2} \frac{1}{\sqrt{2 
E_{p_2} V}} \, e^{- i P_2 \cdot \xi} \\ 
& & \times \sum_{{\bf q}} \frac{1}{\sqrt{2 E_q V}} \, e^{i Q \cdot 
\xi} \sum_{{\bf p}_1'} \frac{1}{\sqrt{2 E_{p_1'} V}} \, e^{i P_1' 
\cdot y'} \\ 
& & \times \int d^{\, 4} \xi' \sum_{{\bf p}_2'} \frac{1}{\sqrt{2 
E_{{p_2'}} V}} \, e^{i P_2' \cdot \xi'} \sum_{{\bf q}'} 
\frac{1}{\sqrt{2 E_{q'} V}} \, e^{- i Q' \cdot \xi'} \\ 
& & \times {\cal S} i \Delta^{(0)} (\xi - x') \left( i \Delta^{(0)} 
(\xi' - y') \right)^* \, , 
\end{eqnarray*} 
where $\Delta^{(0)}$ is the vacuum-theory propagator of $\phi$ and 
${\cal S}$ (cf. Eq.~(\ref{sin1})) takes the form 
\begin{eqnarray*} 
{\cal S} & = & \langle a^\dagger_{{\bf p}_1'} a^\dagger_{{\bf p}_2'} 
a_{{\bf q}'} a^\dagger_{{\bf q}} a_{{\bf p}_1} a_{{\bf p}_2} 
\rangle \\ 
& = & \langle a^\dagger_{{\bf p}_1'} a^\dagger_{{\bf p}_2'} 
(\delta_{{\bf q}', \, {\bf q}} + a^\dagger_{{\bf q}} a_{{\bf 
q}'}) a_{{\bf p}_1} a_{{\bf p}_2} \rangle \, . 
\end{eqnarray*} 
We compute the contributions that include only two-point functions. 
If necessary, the contributions including initial correlations may 
be written down in a straightforward manner. Keeping the terms that 
do not vanish kinematically, we have 
\begin{eqnarray*} 
{\cal S} & = & {\cal S}_1 + {\cal S}_2 \, , \nonumber \\ 
{\cal S}_1 & = & \langle a^\dagger_{{\bf p}_1'} a_{{\bf p}_1} 
\rangle \langle a^\dagger_{{\bf p}_2'} a_{{\bf p}_2} \rangle \, 
[ \delta_{{\bf q}, \, {\bf q}'} + \langle a^\dagger_{{\bf q}} 
a_{{\bf q}'} \rangle ]  \, , \nonumber \\ 
{\cal S}_2 & = & {\cal S}_1 \, \rule[-2.8mm]{.14mm}{7.5mm} 
\raisebox{-2.0mm}{\scriptsize{$\; {\bf p}_1 \leftrightarrow 
{\bf p}_2$}} \, . 
\end{eqnarray*} 
We compute the contribution ${\cal N}_1$ from ${\cal S}_1$. The 
contribution from ${\cal S}_2$ may be computed similarly. Following 
the procedure presented above, we obtain 
\widetext 
\begin{eqnarray*} 
{\cal N}_1 & = & g^2 \lambda^2 \int d^{\, 4} x' G^* (x') \int d^{\, 
4} y' G (y') \int d^{\, 4} \xi \, \int d^{\, 4} \xi' \\ 
&& \times 
\sum {\mbox{\hspace*{-0.525cm}}} \int {\cal D}^{\, 4} P_1 \, 
e^{- i P_1 \cdot (x' - y')} 2 \pi \delta_+ (P_1^2 - m^2) 
N \left( \frac{x' + y'}{2}; P_1 \right) \\ 
&& \times 
\sum {\mbox{\hspace*{-0.525cm}}} \int {\cal D}^{\, 4} P_2 \, 
e^{- i P_2 \cdot (\xi - \xi')} 2 \pi \delta_+ 
(P_2^2 - m^2) N \left( \frac{\xi + \xi'}{2}; P_2 \right) \\ 
&& \times 
\sum {\mbox{\hspace*{-0.525cm}}} \int {\cal D}^{\, 4} Q \, 
e^{- i Q \cdot (\xi' - \xi)} 2 \pi \delta_+ (Q^2 - m^2) \left\{ 1 + 
N \left( \frac{\xi + \xi'}{2}; Q \right) \right\} \\ 
&& \times 
\sum {\mbox{\hspace*{-0.525cm}}} \int {\cal D}^{\, 4} P' \, 
e^{- i P' \cdot (\xi - x')} \frac{i}{P^{' 2} - m^2 + i 0^+} 
\sum {\mbox{\hspace*{-0.525cm}}} \int {\cal D}^{\, 4} Q' \, 
e^{- i Q' \cdot (y' - \xi')} \frac{- i}{Q^{' 2} - m^2 - i 0^+} \, . 
\end{eqnarray*} 
Here we observe that 
\[ 
\frac{\xi + \xi'}{2} - \frac{x' + y'}{2} = \frac{1}{2} \left[ 
(\xi - x') + (\xi' - y') \right] \to \frac{- i}{2} \left( 
\frac{\partial}{\partial P'} - \frac{\partial}{\partial Q'} \right) 
\, , 
\] 
where the partial derivatives are understood to act on the 
\lq\lq propagators'' in momentum representation. 

Making the plane-wave approximation for $G (x)$, 
\[ 
G (x) = \frac{e^{- i R \cdot x}}{\sqrt{2 E_\Phi V}} \;\;\;\;\;\; 
(E_\Phi = \sqrt{r^2 + M^2}) \, , 
\] 
we finally obtain, within the gradient approximation, 
\begin{eqnarray} 
{\cal N}_1 & \simeq & \frac{g^2 \lambda^2}{2 E_\Phi V} \int d^{\, 
4} X \sum {\mbox{\hspace*{-0.525cm}}} \int {\cal D}^{\, 4} P_1 
\sum {\mbox{\hspace*{-0.525cm}}} \int {\cal D}^{\, 4} P_2 \left[ 2 
\pi \delta_+ (P_1^2 - m^2) \, \tilde{N} (X; P_1) \right] \nonumber 
\\ 
&& \times \left[ 2 \pi \delta_+ (P_2^2 - m^2) \, \tilde{N} (X_2; 
P_2) \right] \left[ 2 \pi \delta_+ (Q^2 - m^2) \{ 1 + \tilde{N} 
(X_1; Q) \} \right] \nonumber \\ 
&& \times \left[ 1 - \frac{i}{2} \left( 
\stackrel{\leftarrow}{\partial}_{X_2} + 
\stackrel{\leftarrow}{\partial}_{X_1} \right) \cdot \left( 
\frac{\stackrel{\rightarrow}{\partial}}{\partial P'} - 
\frac{\stackrel{\rightarrow}{\partial}}{\partial Q'} 
\right) \right] \frac{1}{P^{' 2} - m^2 + i 0^+} \nonumber \\ 
&& \times \frac{1}{Q^{' 2} - m^2 - i 0^+} \, 
\rule[-5.0mm]{.14mm}{11.5mm} \raisebox{-4.4mm}{\scriptsize{$\; X_1 
= X_2 = X$, $Q' = P'$}} \, ,  
\label{owae} 
\end{eqnarray} 
\narrowtext 

\noindent 
where $X = (x' + y') / 2$ and $P' = Q' = P_1 - R$ and $Q = P_2 + P_1 
- R$. 

Eq.~(\ref{owae}) corresponds to a contribution to the amplitude in 
the CTP formalism of the \lq\lq process'' (cf.~Eq.~(\ref{AA})), 
$\Phi_2 (R) \to \Phi_1 (R)$, and constitutes a part of the diagram 
as depicted in Fig.~3 in the CTP formalism. As a matter of fact, 
Eq.~(\ref{owae}) represents Fig.~3 with ($p_{1 0} > 0, p_{2 0} > 0, 
q_0 > 0$) plus Fig.~3 with ($p_{1 0} > 0, p_{2 0} < 0, q_0 < 0$). 
 
\begin{figure} 
\caption{A diagram for ${\cal F}_i$ in Eq.~(\ref{owayo}). $i$, $j$, 
$k$ , and $l$ are the vertex-type. Each $\Phi$ is either type-1 or 
type-2. 
\label{f1}} 
\caption{A diagram representing ${\cal N}$, Eq.~(\ref{cal-N}), for 
the process (\ref{reac}). The spacetime points $\xi$ and $x'$ 
($\xi'$ and $y'$) are connected by a vacuum-theory propagator. The 
dot-dashed line stands for the final-state-cut line. The group of 
particles on top of the figure represents the spectator particles. 
\label{f2}} 
\caption{An amplitude for the \lq\lq process'' $\Phi_2 (R) \to 
\Phi_1 (R)$ in the CTP formalism, a part of which represents the 
contribution (\ref{owae}). 
\label{f3}} 
\end{figure} 
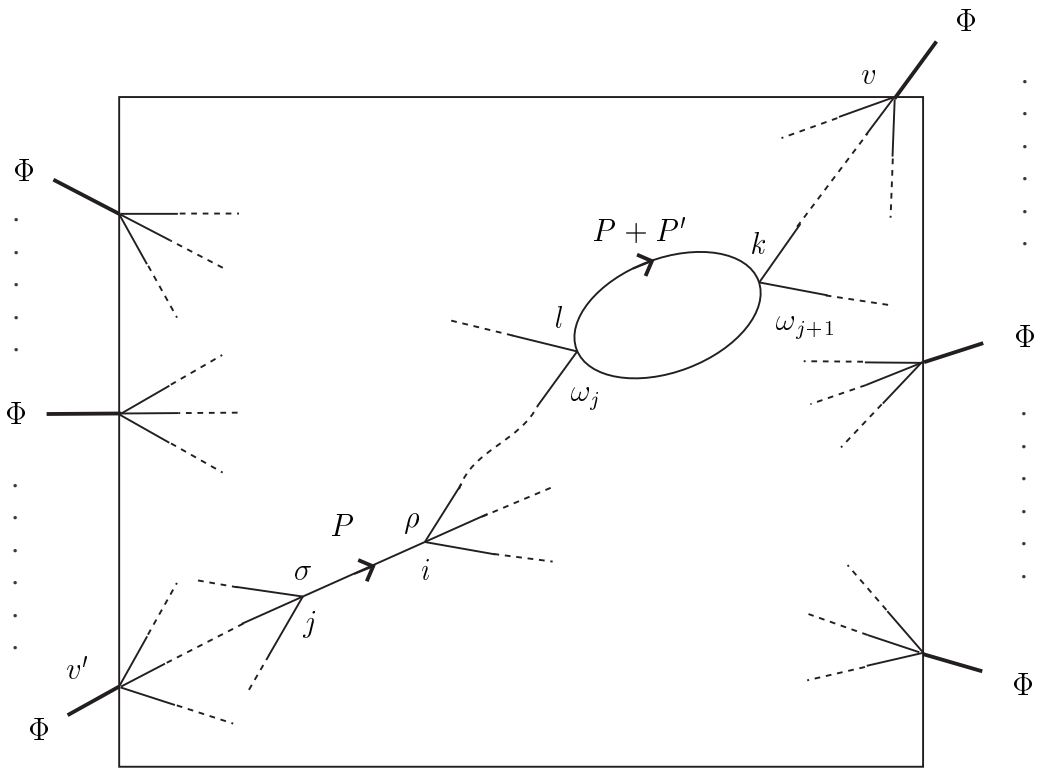
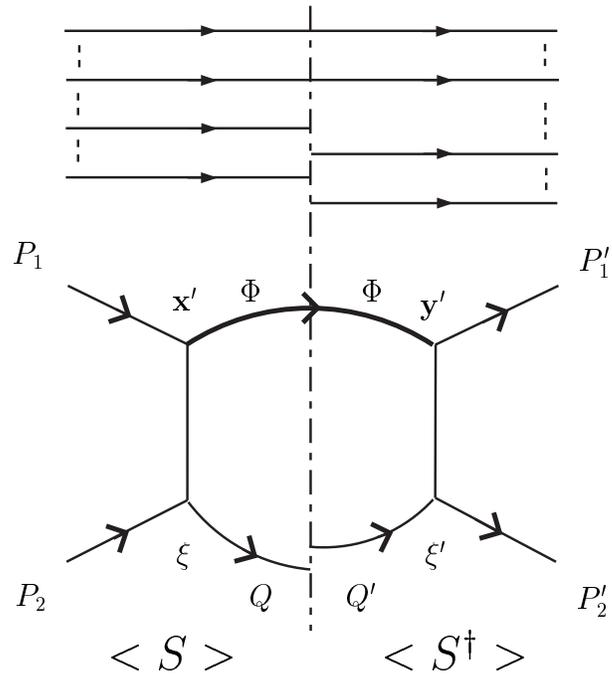
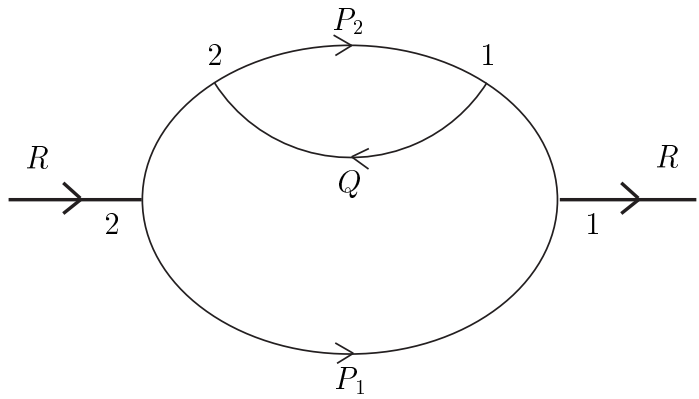
\end{document}